\documentclass[aps,twocolumn,pra,floatfix]{revtex4-1}
\usepackage{placeins}
\usepackage{amsmath,amssymb,bm}
\usepackage{graphicx}
\usepackage{epstopdf}
\usepackage{latexsym}
\usepackage{subfigure}
\usepackage[usenames,dvipsnames]{color}
\usepackage{hyperref}
\usepackage{natbib}
\usepackage{color}
\hypersetup{
  colorlinks,
  citecolor=Blue,
  linkcolor=Red,
  urlcolor=Blue}

\begin{document}
\newcommand{\cs}[1]{{\color{blue}$\clubsuit$#1}}

\title{Dynamics of Vector Solitons in Bose-Einstein Condensates}
\author{Majed O. D. Alotaibi and Lincoln D. Carr}
\affiliation{Department of Physics, Colorado School of Mines, Golden, CO 80401, USA}

\begin{abstract}
We analyze the dynamics of two-component vector solitons, namely bright-in-dark solitons, via the variational approximation in Bose-Einstein condensates. The system is described by a vector nonlinear Schr\"odinger equation appropriate to multi-component Bose-Einstein condensates (BECs). The variational approximation is based on hyperbolic secant (hyperbolic tangent) for the bright (dark) component, which leads to a system of coupled ordinary differential equations for the evolution of the ansatz parameters. We obtain the oscillation dynamics of two-component dark-bright vector solitons. Analytical calculations are performed for same-width components in the vector soliton and numerical calculations extend the results to arbitrary widths. We calculate the binding energy of the system and find it proportional to the intercomponent coupling interaction, and numerically demonstrate the break up or unbinding of a dark-bright soliton. Our calculations explore observable eigenmodes, namely the internal oscillation eigenmode and the Goldstone eigenmode. We find analytically that the density of the bright component is required to be less than the density of the dark component in order to find the internal oscillation eigenmode of the vector soliton and support the existence of the dark-bright soliton. This outcome is confirmed by numerical results. Numerically, we find that the oscillation frequency is amplitude independent. For dark-bright vector solitons in $^{87}$Rb we find that the oscillation frequency range is 90 to 405 Hz, and therefore observable in multi-component BEC experiments.

\end{abstract}
\maketitle

\section{Introduction}
\label{sec:FRHPRA:introduction}

Nonlinear waves have been a fascinating subject since the discovery of the solitary wave in 1834 by John Scott Russell in the Union Canal in Scotland where he observed the “great wave of translation” as he called it~\cite{dauxois2006physics}. Since then, solitary waves of all kinds have been observed in many systems. Solitons in BECs, which are the subject of this paper, have been the focus of research efforts since the creation of BECs~\cite{Kevrekidis2008,Kevrekidis2015}.

A special structure of a coupled dark-bright vector soliton may exist in two-component BECs with repulsive interatomic interactions, where a dark soliton in one component creates a potential well that traps a bright soliton in the second component~\cite{Becker2008b,Busch2001a,Achilleos2012,Hamner2011b,Rajendran2009b,Zhang2009b,Hoefer2011b}. Although a bright soliton does not exist in a system with repulsive interactions~\cite{doi:10.1142/S0217979205032279}, it can be supported in such a binary system due to the nonlinear interaction with the dark soliton component. These solitons can be referred to as symbiotic~\cite{Perez-Garcia2005,Achilleos2012}. A similar possibility of such a mechanism was proposed early in the literature in terms of a Bose-Fermi mixture where bosons and fermions attract each other but the interaction between the bosons themselves is repulsive~\cite{Karpiuk2004a}. Vector solitons also exist in fiber optics~\cite{Kaup1993b, Malomed1991c, Scheuer2001} including bright-bright vector solitons~\cite{Kartashov2011} and dark-bright vector solitons~\cite{Afanasyev1989}. Different types of vector solitons in multiple component BECs, such as pseudo-spinor BECs or three- and higher-component spinor BECs~\cite{Frantzeskakis2010, Kevrekidis2008a}, can be created and transformed into each other by tuning the inter-component interaction via Feshbach resonances~\cite{Zhang2009b, Liu2009b, Qiu-Yan2010}. Examples of these vector solitons in two-component BECs include bright-bright vector solitons~\cite{Xun-Xu2011} and dark-dark vector solitons~\cite{Liu2012, Hoefer2011b} which exhibit rich dynamical far-from-equilibrium phenomena such as beating dark-dark vector solitons~\cite{Cuevas2012}. Among the techniques to create dark-bright solitons in a binary mixture of BECs are phase imprinting~\cite{Becker2008b} and counter-flowing of two binary BEC mixtures~\cite{Middelkamp2011}.

Many studies have been conducted to investigate the oscillation of vector solitons to gain a better understanding of the dynamics of multicomponent nonlinear excitations. The oscillation of two bright vector solitons is one example of these studies. Another example is the oscillation of two dark vector solitons. In the case of dark-bright solitons, there have been many studies to investigate the oscillation of multiple dark-bright vector solitons~\cite{Middelkamp2011,Achilleos2012,Li2009} and the oscillation of the internal modes for two bright vector solitons using a Gaussian ansatz~\cite{Malomed1998} via variational approximation methods. However, to the best of our knowledge no one has treated the internal oscillations of the dark-bright soliton case variationally using hyperbolic functions, which is the subject of this paper. A popular choice for the ansatz in the variational approximation method is Gaussian functions for their relative ease in calculating integrals for the Lagrangian density of two-component vector solitons. In addition, Gaussian functions do not impose any restriction in the choice of the width of the two components in the vector soliton. A disadvantage of using Gaussian functions is that they are less accurate than using hyperbolic functions -- in fact it is exactly the non-Gaussianity of solitons that sets them apart from wavepacket solutions to the linear Schr\"odinger equation. Thus in this paper we perform calculations with variational approximation methods using hyperbolic secant (hyperbolic tangent) for the bright (dark) component in the dark-bright vector soliton. This choice imposes restrictions on the width of the two components such that they must be identical in order to solve the integrals for the Lagrangian density analytically. We study the behavior of the dark-bright  soliton when a phase is imprinted on the bright component only and find the oscillation modes of the system in addition to the binding energy and the velocity of the dark-bright soliton which is effected by the interaction coefficient between the two components. In this scenario the moving bright component pulls the dark component along with it, and oscillates in addition to moving the dark-bright soliton as a whole. One can think of this mode as an vibrational excitation of the dark-bright ``soliton molecule'' as two-component vector solitons are sometimes termed. We will use the term \textit{dark-bright soliton} to describe these vector solitons. Our calculation shows that the system has a second oscillation mode in addition to the vibrational mode, namely a Goldstone mode~\cite{Goldstone2008}, as expected since the whole dark-bright soliton is moving.

The paper is organized as follows. In section~\ref{sec:FRHPRA:The Model} we study oscillation of the two components in the dark-bright soliton when we imprint a phase on the bright component only and find the normal modes of the system by means of variational approximation method based on hyperbolic secant (hyperbolic tangent) for the bright (dark) soliton component for the two-component ansatz. In section~\ref{sec:FRHPRA:Binding Energy of Vector Soliton} we find the binding energy between the bright and dark component in the dark-bright soliton as a function of the distance between the center of each component. In section~\ref{sec:FRHPRA:Numerical Calculations} we investigate the interaction by numerically integrating the dimensionless nonlinear Schr\"odinger equation using an algorithm that is pseudo-spectral in time and adaptive Runge-Kutta in space, between the two components for different interaction coefficients and discuss real experimental values for the internal oscillation in $^{87}$Rb. Finally, we present our conclusions in section~\ref{sec:FRHPRA:Conclusions}.


\section{Analytical Calculations}
\label{sec:FRHPRA:The Model}
\subsection{Lagrangian Density and Ansatz}

The two-component dark-bright soliton is governed by coupled NLSEs~\cite{Kevrekidis2008}, which describe the evolution of the macroscopic wave functions of Bose condensed atoms:
\begin{align}
\label{eq:1DSE}
i \hbar \frac{\partial}{\partial \tilde{t}} \tilde{u}\left(\tilde{x},\tilde{t}\right) &= -\frac{\hbar^2}{2 m}\frac{\partial^2 \tilde{u}\left(\tilde{x},\tilde{t} \right)}{\partial \tilde{x}^2} +\biggl[  \tilde{g}_{11} |\tilde{u}\left(\tilde{x},\tilde{t} \right)|^2
\biggl. \\ \biggl. \nonumber & - \tilde{u}^2_{0} + \tilde{g}_{12} |\tilde{v}\left(\tilde{x},\tilde{t} \right)|^2 \biggl] \tilde{u}\left(\tilde{x},\tilde{t} \right), \nonumber \\ \nonumber
i \hbar \frac{\partial}{\partial \tilde{t}} \tilde{v}\left(\tilde{x},\tilde{t}\right) &= -\frac{\hbar^2}{2 m}\frac{\partial^2 \tilde{v}\left(\tilde{x},\tilde{t} \right)}{\partial \tilde{x}^2} +\biggl[  \tilde{g}_{22} |\tilde{v}\left(\tilde{x},\tilde{t} \right)|^2 \biggl. \\ \biggl. \nonumber & +\tilde{g}_{21} |\tilde{u}\left(\tilde{x},\tilde{t} \right)|^2 \biggl] \tilde{v}\left(\tilde{x},\tilde{t} \right),
\end{align}
%
where tildes denote dimensional quantities. The wave function of the dark soliton is given by $\tilde{u}\left(x,t\right)$  and of the bright soliton by $\tilde{v}\left(x,t\right)$. The interaction strength, $\tilde{g}_{ij} = 2  a_{ij} N \hbar \omega_{\bot}$ for $\left(i,j=1,2\right)$, is renormalized to 1D~\cite{Carr2000a} where $\tilde{g}_{12}$ and $\tilde{g}_{21}$ is the inter-atomic interaction between the two components of the BEC and $\tilde{g}_{11}$ ($\tilde{g}_{22}$) represents the intra-atomic interaction for the dark (bright) component. The dark soliton wave function is rescaled to remove the background contribution, $\tilde{u}_{0}$, as is standard to avoid divergent normalization and energy~\cite{Kivshar1995}. The $s$-wave scattering length between components $i$ and $j$ is $a_{ij}$, $\textit{N}$ is the total number of atoms and  $\omega_{\bot}$ is the oscillation frequency of the transverse trap. We assume the atomic masses for the two components $m_{1}$ and $m_{2}$ are equal to $m$, as appropriate for the case of multiple hyperfine components of e.g. $^{87}$Rb. To nondimensionlize Eqs.~\eqref{eq:1DSE} we multiply them by $\left(\hbar \omega_{\bot}\right)^{-1}$ and scale all quantities according to the following units:


\begin{equation}
	\label{eq:scaling_units}
\begin{aligned}
x & = \frac{\tilde{x}}{\ell_{\bot}}, \\
t & =  \tilde{t} \omega_{\bot}, \\
g_{ij} & = \frac{\tilde{g}_{ij}}{\ell_{\bot} \hbar \omega_{\bot}},  \\
|u|^2 & = \ell_{\bot} |\tilde{u}|^2 ,\\
|v|^2 & = \ell_{\bot} |\tilde{v}|^2 ,\\
u^2_{0} & = \frac{\tilde{u}^2_{0}}{\hbar \omega_{\bot} } ,
\end{aligned}
\end{equation}
where $\ell_{\bot}=\sqrt{\hbar /\left(m \omega_{\bot}\right)}$ is the harmonic oscillator length. In Section~\ref{sec:FRHPRA:Units} we discuss specific choices that are consistent with experimental observations. For simplicity we take $g_{11}\equiv g_{1}, g_{22}\equiv g_{2}, g_{12}$ and $g_{21}\equiv g$. The dimensionless NLSE becomes

\begin{align}
	\label{eq:coupledNLSE}
i \frac{\partial u}{\partial t} &= -\frac{1}{2} \frac{\partial^2 u}{\partial x^2} + \left[g_{1} \left| u \right|^2 -u^2_{0} + g \left| v \right|^2 \right] u, \nonumber \\
i \frac{\partial v}{\partial t} &= -\frac{1}{2} \frac{\partial^2 v}{\partial x^2} + \left[g_{2} \left| v \right|^2  + g \left| u \right|^2 \right] v.
\end{align}

We work with the dimensionless 1D two-component coupled NLSE, Eq.~\eqref{eq:coupledNLSE}, throughout the rest of the paper. We use the normalization conditions
\begin{subequations}
	\label{eq:normalizations}
	\begin{align}
	\int_{-\infty}^{\infty} dx\; \left(\frac{u^2_{0}}{g_{1}} - \left|u\right|^2\right) =\frac{N_{1}}{N}, \\
	\int_{-\infty}^{\infty} dx\; \left| v\right|^2 = \frac{N_{2}}{N},
\end{align}
\end{subequations}
for the dark and bright component, respectively, and take the total number of atoms to be constant,
\begin{equation}
	\label{eq:total number of atmos}
	N_{1} + N_{2}=N.
\end{equation}
This choice of normalization includes the actual number of atoms, $\textit{N}$, in the definition of the nonlinear coefficient $\tilde{g}_{ij}$~\cite{Kevrekidis2008}. To obtain Eq.~\eqref{eq:coupledNLSE}, we introduce the following Lagrangian density where we use Euler-Lagrangian equations to get the equations of motion, i.e., the coupled NLSE of Eq.~\eqref{eq:coupledNLSE}:

	\begin{align}
	\label{eq:LagrangiandDensity}
\begin{split}
{\mathcal{L}} = & \frac{i}{2} \left[u^* \frac{\partial u}{\partial t} -u \frac{\partial u^*}{\partial t}  \right] \left[1-\frac{u^2_{0}}{g_{1} \left| u \right|^2} \right]
- \frac{1}{2} \left|  \frac{\partial u }{\partial x}  \right|^2
\\ &
- \frac{1}{2} \left[\sqrt{g_{1}} \left|u\right|^2 - \frac{u^2_{0}}{\sqrt{g_{1}}}  \right]^2
+ \frac{i}{2} \left[v^* \frac{\partial v}{\partial t} -v \frac{\partial v^*}{\partial t}  \right] \\ &
- \frac{1}{2} \left|  \frac{\partial v }{\partial x}  \right|^2
-\frac{g_{2}}{2} \left| v \right|^4  - g \left| u \right|^2 \left| v \right|^2 \\&
+\frac{u^2_{0}}{2g_{1}}\left[2 \theta_{2} \left(x+ d\left(t\right) \right) +\theta_{1}\left(t\right) \right]^2.
\end{split}
\end{align}
%
Note that the last term does not depend on the wave function of the dark or the bright component and it was added to eliminate the infinity when using the ansatz, Eq.~\eqref{eq:ansatz}, with $\theta_1$ and $\theta_2$ to be defined in the following.
We adopt the following trial functions as the dark-bright soliton solutions to Eq.~\eqref{eq:coupledNLSE}:
%
\begin{align}
\label{eq:ansatz}
u\left(x,t\right) &= \frac{u_{0}}{\sqrt{g_{1}}} \left\{ i A + c\ \mathrm{tanh}\left[\frac{\left(d\left(t\right) + x \right)}{w} \right] \right\}   \\ \nonumber & \times \mathrm{exp}{\left\{i \left[\theta_{0} + \left(d\left(t\right) + x \right) \theta_{1}\left(t\right)+ \left(d\left(t\right) + x \right)^2 \theta_{2} \right]\right\}}, \nonumber \\
v\left(x,t\right) &= \frac{u_{0}}{\sqrt{g_{2}}} F \ \mathrm{sech} \left[\frac{\left(b\left(t\right) + x \right)}{w} \right] \nonumber
\\  \nonumber & \times  \mathrm{exp}{\left\{i \left[\phi_{0} + \left(b\left(t\right) + x \right) \phi_{1}\left(t\right)+ \left(b\left(t\right) + x \right)^2 \phi_{2} \right]\right\}}.
\end{align}
The parameters $A$, $c$ and $F$ describe the amplitude of the two components, where $A^2+c^2=1$, as is standard in the formulation of an NLSE dark soliton~\cite{Kivshar1998}. In the exponential terms, $\phi_{0}$ and $\theta_{0}$ give rise to a complex amplitude. $\phi_{1}\left(t\right)$ and $\theta_{1}\left(t\right)$ are responsible for the dark and bright component velocities. Note that the velocity of a dark soliton also depends on the amplitude of the wave function as shown in Eq.~\eqref{eq:EvolutionEq4b}, $\phi_{2}$ and $\theta_{2}$ are essential to vary the width~\cite{Ueda1990}, and $d\left(t\right)$ and  $b\left(t\right)$ are the position of the dark and bright soliton, respectively. The two components have the same width $w$. To study the oscillation of the two components in time, we chose the variational parameters to be the two component positions $d\left(t\right)$ and $b\left(t\right)$ and the phases $\theta_{1}\left(t\right)$ and $\phi_{1}\left(t\right)$. As mentioned in the introduction, the analytical calculations using hyperbolic functions as an ansatz, which are more accurate than using Gaussian functions, will impose a limit on the choice of the width of the two components where we have to choose an identical width for the two components as opposed to using a Gaussian ansatz~\cite{Malomed1998}. However, we will relax this constraint in our numerical studies.


\subsection{Evolution Equations}
Substituting Eq.~\eqref{eq:ansatz} into the Lagrangian density Eq.~\eqref{eq:LagrangiandDensity} and integrating over space from $-\infty$ to $\infty$, results in the Lagrangian as a function of the variational parameters. Then, using the Euler-Lagrange equation gives a system of ordinary differential equations (ODEs) that describes the evolution in time for the position and phase of both components:



\begin{subequations}
	\label{eq:EvolutionEqAll}
\begin{align}
&\frac{d}{dt} b\left(t\right) =  -\phi_{1}\left(t \right), \\
\label{eq:EvolutionEq4b}
&\frac{d}{dt} d\left(t\right) =  - \alpha - \theta_{1} \left(t\right), \\
&\frac{d}{dt} \theta_{1}(t) = \beta \frac{d}{dt} \phi_{1}\left(t\right), \\
\label{eq:8_d}
&\frac{d}{dt} \phi_{1}(t) = \gamma \, \mathrm{csch}\left(\frac{b\left(t\right) - d\left(t\right) }{w}\right)^4 \left\{2  \left(b\left(t\right) - d\left(t\right)\right) \right.  \\ \nonumber  & \left. \times
\left[ 2 +\mathrm{cosh}\left( 2\frac{b\left(t\right) - d\left(t\right) }{w} \right) \right] - 3w\,\mathrm{sinh}\left( 2\frac{b\left(t\right) - d\left(t\right) }{w} \right) \right\}.
\end{align}
\end{subequations}
Where,
\begin{equation}
	\label{constants}
	\alpha = \frac{A}{c w}, \;
	\beta =\frac{F^2 g_{1}}{c^2 g_{2}}, \;
	\gamma = \frac{ c^2 g u_{0}^2}{g_{1} w^2}.
\end{equation}


Equations.~\eqref{eq:EvolutionEqAll} can be reduced to one second order ODE:

\begin{equation}
\begin{aligned}
	\label{eq:2ndODE}
\frac{d^{2}}{dt^{2}}{l}(t) =& (\beta-1)\gamma \;\mathrm{csch} (\frac{l(t)}{w})^4 \left \{ 2 l(t) \left [ 2 + \mathrm{cosh}(\frac{2 l(t) }{w})\right ] \right. \\& \left.  - 3w \;\mathrm{sinh}\,(\frac{2 l(t) }{w}) \right\}.
\end{aligned}
\end{equation}
where $l(t)=b(t)-d(t)$. Despite the attractive simplicity of this unified description, it is physically advantageous to address the problem with Eqs.~\eqref{eq:EvolutionEqAll} to illustrate the behavior of the evolution of the variational parameters in time and to clarify the physical meaning of the fixed point and linear stability analysis in the next section.


\subsection{Normal Modes}
\label{sec:FRHPRA:Normal Modes}
\begin{figure}[!htbp]
\centerline{\includegraphics[width=0.8\columnwidth]{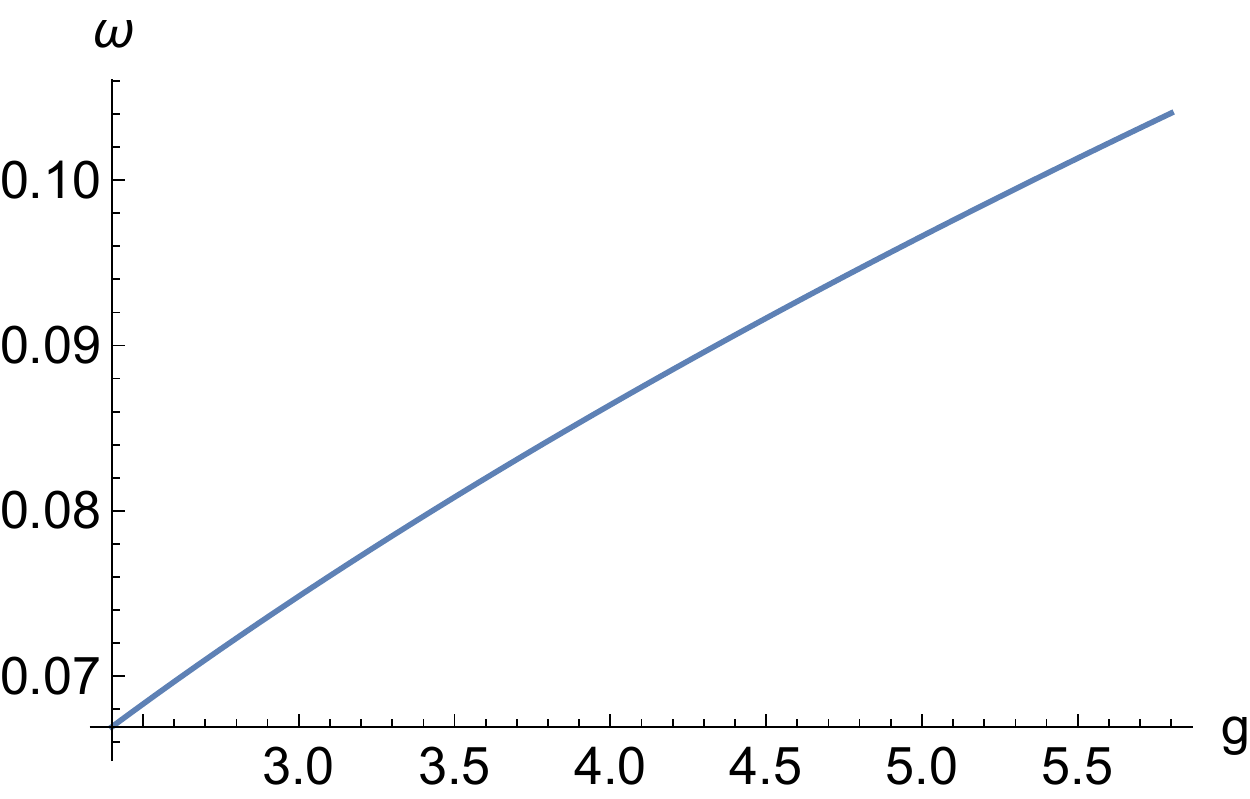}}
\caption{\label{fig:FRHPRA:g_vs_freq_analytical}  \emph{Oscillation frequency of the two components in the dark-bright soliton versus the interaction coefficients, g}. We set $N_{1}\approx0.503*10^5$ atoms and width w=1 where $\omega$ and g are unitless. The range of the values in g is from 2.4 to 5.8 in association with the range of g in the numerical calculations.}
\end{figure}

Equations~\eqref{eq:EvolutionEqAll} possess one stable fixed point:
\begin{equation}
\begin{aligned}
	\label{eq:fixedpoints}
		\phi_{1} =0 ,\ \theta_{1} = - \alpha ,\ l=0.
\end{aligned}
\end{equation}
Since $l=0$, we can choose $b=d=0$. In Appendix~\ref{appendix:singularity} we prove that Eqs.~\eqref{eq:EvolutionEqAll} with the fixed point $l=0$ do not possess singularity. We proceed by linearizing Eqs.~\eqref{eq:EvolutionEqAll} around the fixed point Eq.~\eqref{eq:fixedpoints}, (i.e. $a_{i}\left(t\right) \rightarrow a_{fp} +\delta a e^{\mathrm{i}\omega t}$ ) where $a_{i}$ represents the variational parameters and $a_{fp}$ is the fixed point mentioned above. This result in forming a matrix of the ODEs.

 \begin{equation}
	\label{eq:matrix1}
\begin{bmatrix}
    i\omega & 0 & 1 & 0   \\
    0 & i\omega & 0 & 1 \\
    A_{1} & A_{2} & i\omega & 0 \\
	A_{3} & A_{4} & 0 & i\omega
\end{bmatrix}
\begin{bmatrix}
    \delta b   \\
    \delta d   \\
    \delta \phi_{1} \\
	\delta \theta{1}
\end{bmatrix}
=\begin{bmatrix}
    0   \\
    0   \\
    0	\\
	0
\end{bmatrix}
\end{equation}
where
\begin{align}
A_{1}&= -A_{2} = -\left(8 c^2 g u^2_{0})/\right(15 g_{1} w^2),\\
A_{3}&= -A_{4} = -\left(8 F^2 g u^2_{0})/\right(15 g_{2} w^2).
\end{align}
Solving the determinant of the matrix yields the eigenmodes of the system with eigenvectors
 \begin{subequations}
	\label{eq:eigevectors}	
	\begin{align}	
	&\nu_{\pm}=
	\begin{bmatrix}
    	\pm\frac{1}{2} \sqrt{15}\left(N_{1}/N_{2}\right)\left(w^\frac{3}{2}/\sqrt{g}\right)\sqrt{\frac{N}{N_{2}-N{1}}}  \\
    	\pm\frac{1}{2} \sqrt{15}\left(w^\frac{3}{2}/\sqrt{g}\right)\sqrt{\frac{N}{N_{2}-N{1}}}   \\
    	{N_{1}}/{N_{2}} \\
		1
	\end{bmatrix}, \\
	& \;\;\;\;\;\;\;\;\;\;\;\;\;\;\;\;\;\;\;\;\;\;\;\;\;\;\;\;\;\;\;
	\label{eq:goldstone vector mode}
	\nu_{0}=
	\begin{bmatrix}
		1   \\
		1   \\
 	    0 \\
		0
	\end{bmatrix},
	\end{align}
\end{subequations}
%
and eigenfrequencies \footnote{We note that the simplification of Eq.~\eqref{eq:2ndODE} produces the same eigenfrequencies, as we verified in an independent calculations.}
 \begin{subequations}
	\label{eq:eigenfrequencies}	
	\begin{align}	
	\label{eq:frequency mode}
	&\omega_{\pm}=\pm 2 i \sqrt{\frac{1}{15}}\sqrt{g} w^{-\frac{3}{2}} \sqrt{\left(N_{2}-N_{1}\right)/N}, \\
	\label{eq:goldstone mode}	
	&\omega_{0} = 0,
	\end{align}
\end{subequations}
where the oscillation frequency Eq.~\eqref{eq:goldstone mode} corresponds to the zero-energy mode, sometimes defined in the literature as a Goldstone mode~\cite{Goldstone2008,Roy2014a}. This mode breaks translational symmetry with no energy cost. We can interpret it as moving dark-bright soliton without internal oscillation of the two components. Also, the eigenvector of this mode Eq.~\eqref{eq:goldstone vector mode} shows no contribution from the phases that is responsible in the first place for the oscillation. In Eq.~\eqref{eq:frequency mode}, for stable oscillation we have to meet the condition $N_{1} > N_{2}$ in other words, $\ g_{2} > \frac{F^2}{c^2} g_{1}$ which means for the same amplitude components there is no oscillation. This result is supported by the numerical calculations in section~\ref{sec:Bright in Dark Vector Soliton with equal Interaction Coefficients} where we find that the bright component in dark-bright soliton does not exist when the density of the bright component is equal to or greater than the density of the dark component (Fig.~\ref{fig:FRHPRA:GS_manakov}). Using $N_{2}=N-N_{1}$ we can rewrite the oscillation frequency as
	\begin{equation}
		\begin{aligned}
		\label{eq:oscillationfrequency}
			\omega_{\pm}=\mp 2 \sqrt{\frac{1}{15}} w^{-\frac{3}{2}}\sqrt{g} \sqrt{\left(2N_{1}/N\right) -1}.
		\end{aligned}
	\end{equation}
Note that for a real oscillation the normalization constant $2N_{1}/N$ should be greater than one, which in turn makes $N_{1}>N{_2}$, i.e., the density of the dark component is greater that the density of the bright component. Considering the typical number of atoms in $^{87}$Rb experiment,  we set $N=10^5$ and $N_{1}\approx0.503 \times 10^5$. Setting w = 1 in Eq.~\eqref{eq:oscillationfrequency} we plot the relative frequency versus the interaction coefficient g in Fig.~\ref{fig:FRHPRA:g_vs_freq_analytical}.

\subsection{Binding Energy of Vector Soliton}
\label{sec:FRHPRA:Binding Energy of Vector Soliton}

In the Lagrangian density, Eq.~\eqref{eq:LagrangiandDensity}, the term $g\left| u \right|^2 \left| v \right|^2 $  represents the coupling interaction per unit space between the two components of the dark-bright soliton. Using the ansatz Eq.~\eqref{eq:ansatz}, we can integrate this term over x to find the coupling interaction of the system. The binding energy can be found when we subtract the coupling interaction energy at $l=0$ from $l = \infty$ where $l$ is the separation between the bright and dark solitons.  The energies associated with all other terms in the Lagrangian density turn out to be independent of $l$.  The coupling interaction energy of the system is
\begin{equation}
	\label{eq:coupling energy}
\begin{split}
\mathrm{E_{coupling}}
& = \frac{F^2 g u^4_{0} }{g_{1}g_{2}} \mathrm{csch} \left[\frac{l\left(t\right)}{w}\right]^2 \times \\
&\left[4 c^2 \left(w-l \,\mathrm{coth} \left[\frac{l\left(t\right)}{w}\right] \right) + 2 w  \,\mathrm{sinh}\left[\frac{l\left(t\right)}{w}\right]^2 \right].
\end{split}
\end{equation}
\begin{figure}[!htbp]
\centerline{\includegraphics[width=0.9\columnwidth]{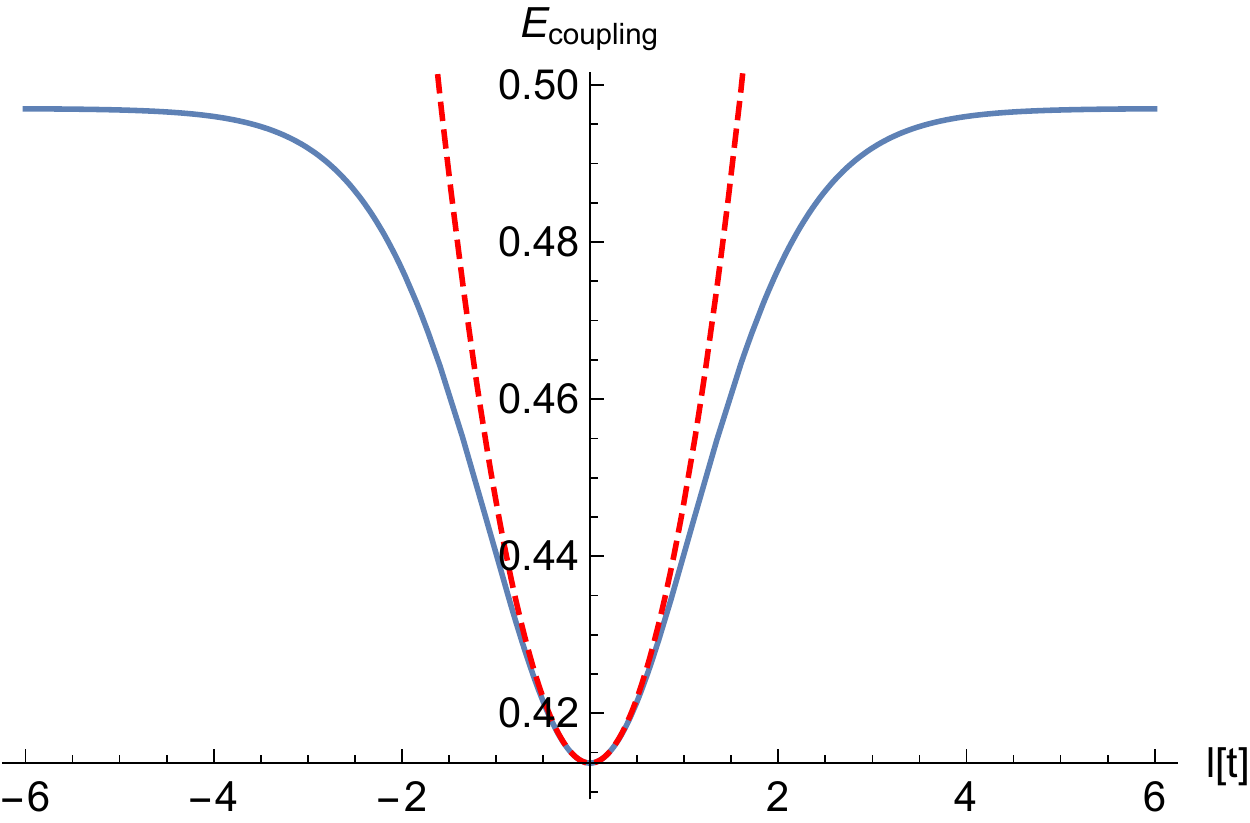}}
\caption{\label{fig:FRHPRA:bindingEnergy} \emph{Coupling energy versus the distance between the two component, $l\left(t\right)$, when t=0}. Here we normalize the interaction coefficients to unity and set $N_{1}\approx0.503*10^5$ atoms. The solid blue line represent Eq.~\eqref{eq:coupling energy} and the dashed red line represent Eq.~\eqref{eq:expansion_coupling_energy}.}
\end{figure}
In Fig.~\ref{fig:FRHPRA:bindingEnergy} we plot Eq.~(\ref{eq:coupling energy}).  As expected for a binding energy, the coupling interaction energy is minimum at the center when the two component locations coincide.  By applying a phase on the bright component it will experience a force due to the coupling interaction energy that will bring it back to the location of the minimum energy which will establish the oscillation of the two components. If the phase imprinted is large enough, the bright soliton will reach a point where the distance between the two components is large and the change in the coupling interaction energy as a function of space is small enough that there will only be a negligible amount of force exerted on the bright component to bring it back to the original location which will break the dark-bright soliton at this point.

To explore the behavior of the oscillation around the fixed point when $l<<1$ we expand Eq.~(\ref{eq:coupling energy}):
\begin{align}	
 \label{eq:expansion_coupling_energy}
\mathrm{E_{coupling}} = \frac{2(3-2c^2) F^2 g u^4_{0} w}{3 g_{1} g_{2}} + \frac{8 c^2 F^2 g u^4_{0}}{15 g_{1} g_{2} w} l^2
\end{align}

As a result, we see that the coupling energy when $l<<1$ behaves as a parabolic potential energy near the fixed point. Therefore, we should expect the oscillation frequency to be an amplitude independent for small amplitude excitations, and this is the result we obtained in the numerical section (see Fig.~\ref{fig:FRHPRA:g_vs_freq}).

We can treat the coupling energy as a potential energy and derive the equation of motion for $l(t)$.
\begin{equation}
	\label{eq:EOMs_from_binding_energy}
	\begin{split}
	m \frac{d^2}{dt^2} l(t) &= -\frac{d}{d l} \mathrm{E_{coupling}} \\
	&= \frac{2 c^2 F^2 g u^4_{0}}{g_{1} g_{2} w} \,\;\mathrm{csch} (\frac{l(t)}{w})^4 \left \{ 2 l(t) \left [ 2 \right.\right. \\& 
\left.\left. + \mathrm{cosh}(\frac{2 l(t) }{w})\right ]   - 3w \;\mathrm{sinh}\,(\frac{2 l(t) }{w}) \right\},
	\end{split}
\end{equation}
where $m=1$ in our units. Comparing Eq.~\eqref{eq:EOMs_from_binding_energy} to Eq.~\eqref{eq:2ndODE} we find that the two equations are different only by the coefficients and therefore yield different frequencies. This can be understood by examining the Lagrangian density, Eq.~\eqref{eq:LagrangiandDensity}, where we subtract the background contributions from the dark soliton momentum term and the intra-component mean field energy term. Thus, The calculations leading to Eq.~\eqref{eq:2ndODE} account for this subtraction whereas the calculations leading to Eq.~\eqref{eq:EOMs_from_binding_energy} is not. Consequently, The coefficients are different.

By taking the difference between Eq.~(\ref{eq:coupling energy}) at $l=0$ and $l=\infty$ we find the binding energy:
\begin{equation}
	\label{eq:binding energy}
\begin{split}
\mathrm{E_{binding}} & = \mathrm{E_{coupling (\textit{l}\to 0)}} - \mathrm{E_{coupling (\textit{l}\to\infty)}} \\
& = -\frac{4 c^2 F^2 g u_{0}^4 w}{3 g_{1} g_{2}} .
\end{split}
\end{equation}
We note that the binding energy is thus proportional to the intercomponent coupling $g$ and inversely proportional to the intracomponent couplings $g_1$, $g_2$.  The latter inverse proportionality is due to normalization. In addition, we calculate the kinetic energy (KE) and the intra-component mean-field energy (MFE) of the dark and bright component, separately, and compare them to the binding energy above.

For the dark component in the dark-bright soliton,
\begin{equation}
	\label{eq:KEDS}
\mathrm{KE} = \frac{c^2 u_{0}^2 [-2+\pi^2 w^4 \theta_{2}^2]}{3 w g_{1}} + \frac{c u_{0}^2 \theta_{1}(t)[2 A + c w \theta_{1}(t)]}{g_{1}},
\end{equation}
\begin{equation}
	\label{eq:MFEDS}
\mathrm{MFE} = -\frac{2 c^4 u_{0}^4 w}{3 g_{1}}.
\end{equation}

For the bright component in the dark-bright soliton,
\begin{equation}
	\label{eq:KEBS}
\mathrm{KE} = -\frac{F^2 u_{0}^2 [1+\pi^2 w^4 \phi_{2}^2]}{3 w g_{2}} - \frac{F^2 u_{0}^2 w \phi_{1}(t)^2}{g_{2}},
\end{equation}
\begin{equation}
	\label{eq:MFEBS}
\mathrm{MFE} = - \frac{2 F^4 u_{0}^4 w}{3 g_{2}}.
\end{equation}
We found the KE and the MFE of the dark (bright) soliton component is inversely proportional to the intracomponent coupling $g_{1}$ ($g_{2}$). Note that both the KE and the MFE of the two components does not depend on the intercomponent coupling $g$ as expected. This result can be understood when we examine the Lagrangian density, Eq.~\eqref{eq:LagrangiandDensity}, where the intercomponent coupling $g$ only appears in the coupling term and therefore it only contributes to the binding energy.

Finally, we compare the binding energy to the kinetic energies (i.e. Eq.~\eqref{eq:KEDS}, Eq.~\eqref{eq:KEBS}) and the mean field energies (i.e. Eq.~\eqref{eq:MFEDS}, Eq.~\eqref{eq:MFEBS}) of the dark-bright vector soliton. We find that in order to break or unbind the dark-bright vector soliton the imprinted phase on the bright component should be greater than the following quantity:

\begin{equation}
	\label{eq:phi_1}
	\begin{split}
	\mathrm{\phi_1} >  \,\, &\frac{1}{\sqrt{3 N_{2}}} [-\frac{2 N_{1}+ N_{2} - 2 N_{1} w}{w^2}-\frac{N^2_{1} (2+g_{1})+g_{2} N^2_{2}}{N w} \\
	 & + \pi^2 w^2(N_{1} \theta^2_{2} - N_{2} \phi^2_{2}) +3 N_{1} \theta^2_{1}  \\
	 & +\frac{6 \,\theta_{1}\sqrt{N_{1}} \sqrt{2 N u^2_{0} w- g_{1} N_{1}}}{w \sqrt{g_{1}}}  ]^{-1}
	\end{split}
\end{equation}

In Section~\ref{sec:Bright in Dark Vector Soliton with Unequal Interaction Coefficients}, we compare Eq.~\eqref{eq:phi_1} to Fig.~\ref{fig:FRHPRA:breaking_VS}.



\section{Numerical Calculations}
	\label{sec:FRHPRA:Numerical Calculations}
	
In this section we investigate the interaction numerically between the two components, first with equal interaction coefficients where we attempt to find the ground state energy of a dark-bright soliton. A criterion that should be met in order to find such a ground state prevents creation of the ground state with equal interaction coefficients. This criterion is that the density of the bright soliton component should be less than the density of the dark component soliton in the dark-bright soliton. In Sec.~\ref{sec:FRHPRA:Normal Modes} we encounter this condition as a requirement to find a real oscillation of the two component dark-bright soliton. Second, we investigate the interaction
between the two components with unequal interaction coefficients by finding the ground state of the system when the interatomic interaction goes from the miscible to the immiscible domain which represents a quantum phase transition for the dark-bright soliton. We calculate the amplitude of the two components, the velocity of the dark-bright soliton and the oscillation frequency mode as a function of the interaction coefficients. Third, we end this section with a discussion of the experimental case for $^{87}$Rb where we can use these units to convert between the dimensionless variables in the study conducted and a physically measurable quantities such as the oscillation time.

\subsection{Bright in Dark Vector Soliton with Equal Interaction Coefficients}
\label{sec:Bright in Dark Vector Soliton with equal Interaction Coefficients}

\begin{figure}[!htbp]
\centerline{\includegraphics[width=\columnwidth]{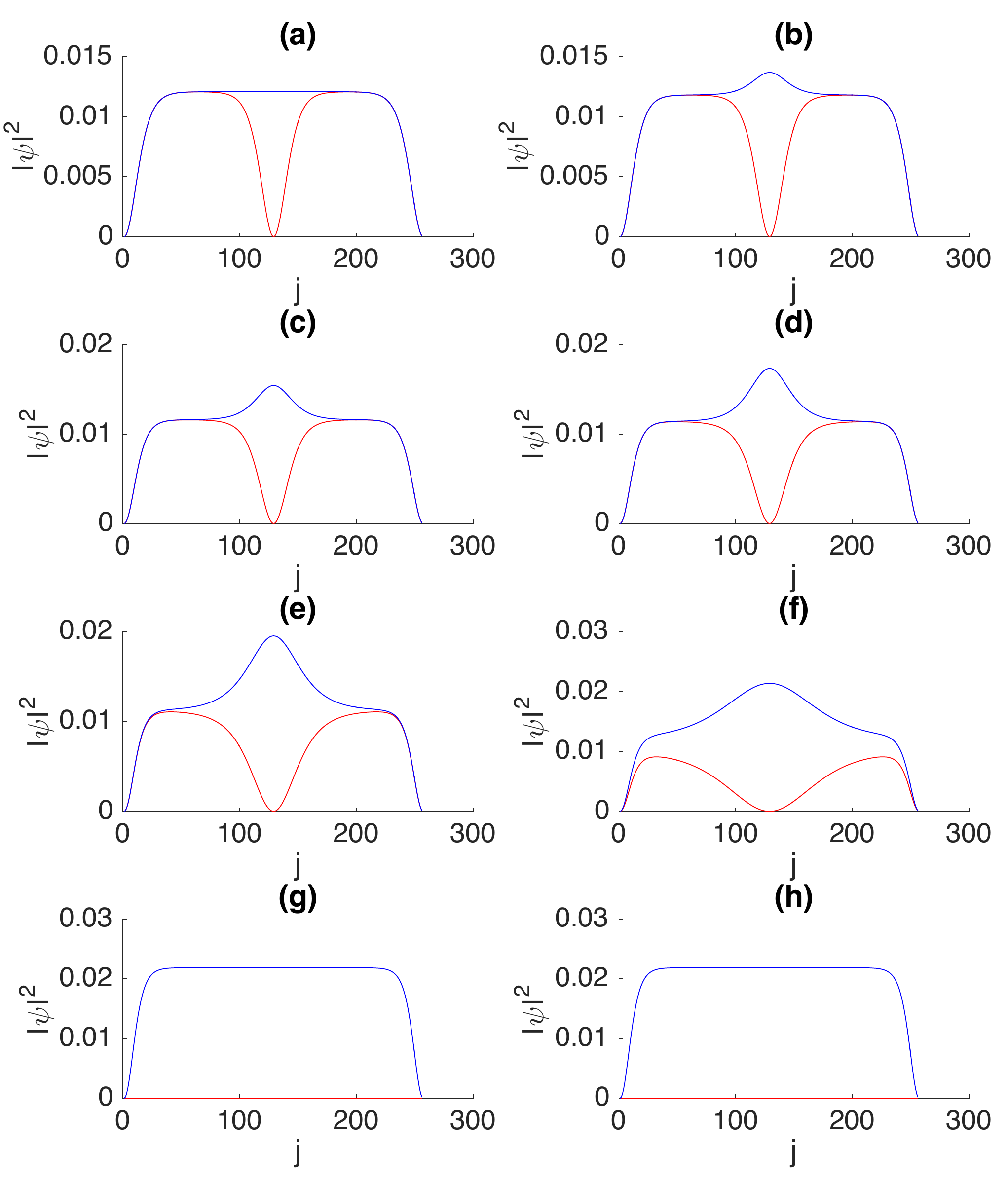}}
\caption{\label{fig:FRHPRA:GS_manakov}  \emph{Ground state density of the dark-bright soliton, when the interaction coefficients are equal to unity, versus the coupling interaction cofficient.} We found the ground state energy of the DB soliton using imaginary time propagation method. The box dimension is L=50. The grid point $j\in \{1,..,n_x\}$ where $n_{x}=256$ is the grid size. In (a)-(h) the values of $g= 0.0, 0.2, 0.4, 0.6, 0.8, 0.95, 1.0, 1.2$, respectively. Note that the bright component takes the form of constant background as we get close to the Manakov case where the dark component is eliminated and what is left is a constant wave function, as is the case for a bright soliton in repulsive media.}
\end{figure}

We start the numerical calculations by using the imaginary-time-propagation method to find the ground state energy of the coupled NLSEs. Starting with constant initial wavefunctions for both components, where we imprinted a phase on the constant dark component only, we find that the ground state energy for the dark-bright soliton exists when the density of the bright component is less than the density of the dark component (i.e., $g_{2}>g_{1}$) see Fig.~\ref{fig:FRHPRA:GS_many}. When the density of both components is equal to each other we do not find that the bright component exists inside the dark soliton as shown in Fig.~\ref{fig:FRHPRA:GS_manakov}. When we reach the Manakov case, $g=g_{1}=g_{2}$, we find that the bright-in-dark soliton does not exist using the initial wave function mentioned above in the imaginary-time-propagation method.

\subsection{Bright in Dark Vector Soliton with Unequal Interaction Coefficients}
\label{sec:Bright in Dark Vector Soliton with Unequal Interaction Coefficients}

\begin{figure}[!htbp]
\centerline{\includegraphics[width=\columnwidth]{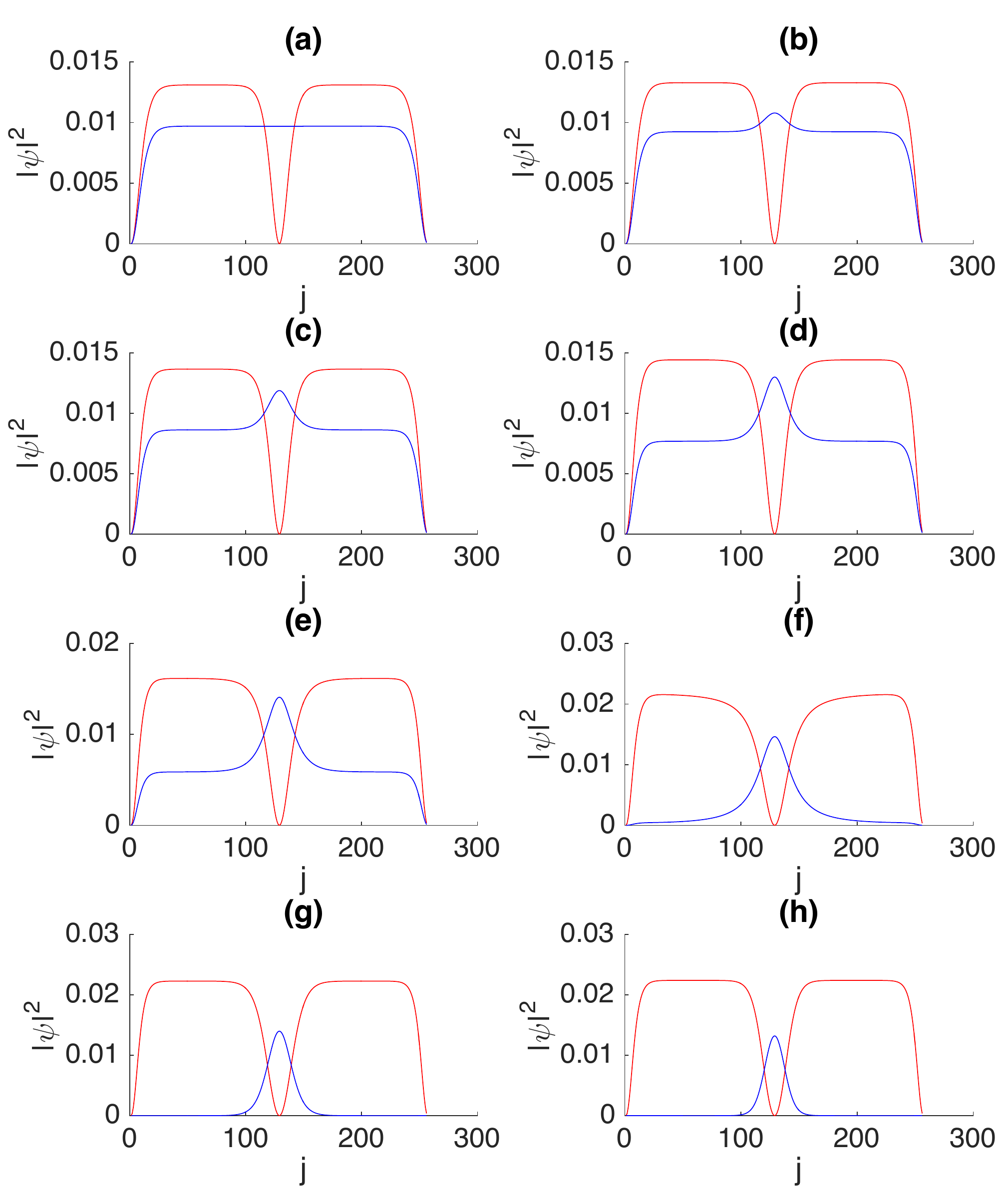}}
\caption{\label{fig:FRHPRA:GS_many}   \emph{Ground state density of the dark-bright soliton in the miscible/immiscible domain with $g_{1}=2.0$ and $g_{2}=2.7$.} Same grid choices as Fig.~\ref{fig:FRHPRA:GS_manakov}. In (a)-(h) the values of $g= 0.0, 0.4, 0.8, 1.2, 1.6, 2.0, 2.4, 2.8$, respectively. Note the phase transition when $g>2.3$ which characterizes the immiscible domain.}
\end{figure}

In the miscible domain in Fig.~\ref{fig:FRHPRA:GS_many}, the strength of the repulsive interaction between the two components is less than the repulsive interaction between the particles in the bright component which allows the bright soliton to expand and reach the boundaries. In the immiscible domain, the coupling interaction is strong to the point that it forces the bright component to live within the dark soliton only.

\begin{figure}[!htbp]
\centerline{\includegraphics[width=\columnwidth]{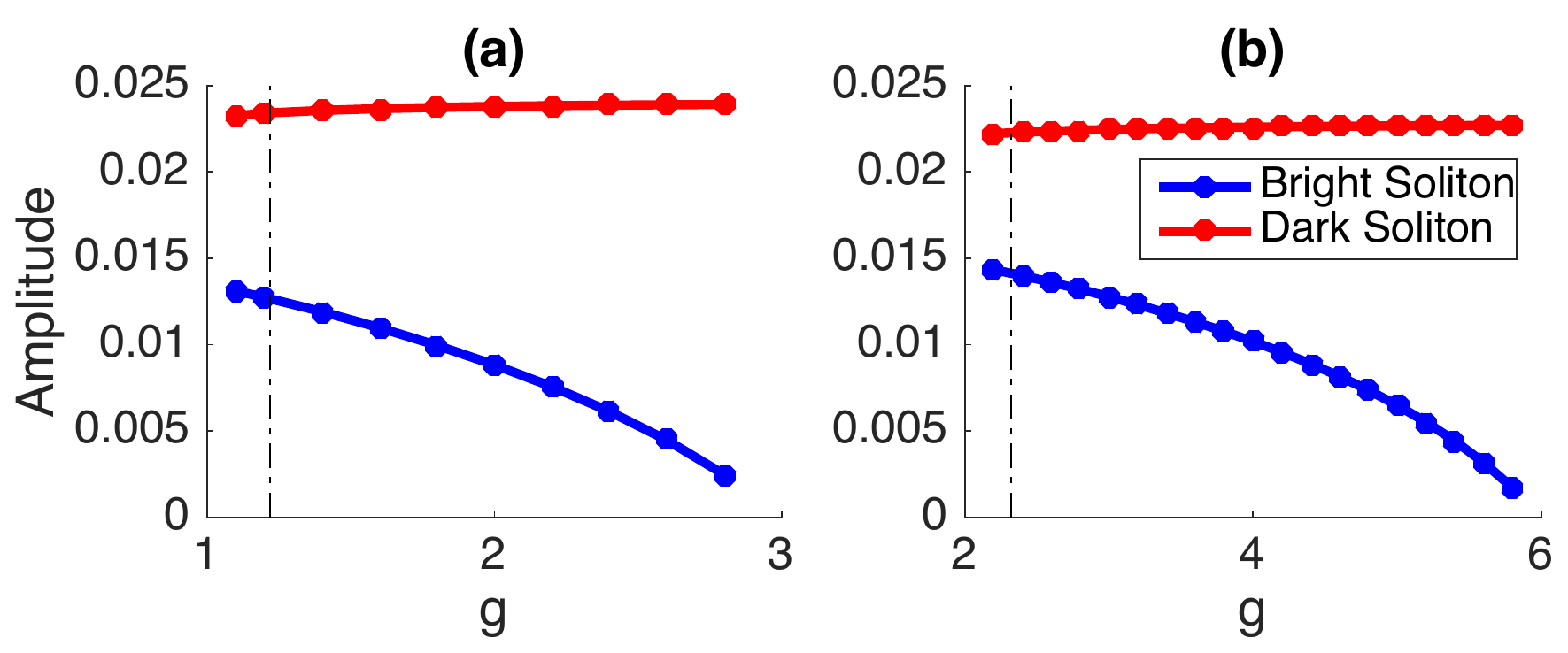}}
\caption{\label{fig:FRHPRA:g_vs_A}   \emph{Amplitude of the bright and dark component versus the coupling interaction g.} We measure the amplitudes of the two components at the ground state with different values of $g_{1}$, $g_{2}$ and $g$. $(a) g_{1}=1.0, g_{2}=1.5. (b) g_{1}=2.0, g_{2}=2.7$.}
\end{figure}


In Fig.~\ref{fig:FRHPRA:g_vs_A}, as we increase the intercomponent coupling, $g$, the amplitude of the bright component decreases and the amplitude of the dark component increases. With increasing the intercomponent coupling, the ground state of the dark-bright soliton shows that the density of the bright component decreases and therefore the amplitude too. This can be understood by examining Fig.~\ref{fig:FRHPRA:GS_many}. We see that when the intercomponent coupling is zero the size of the two densities of the dark and bright component is governed by the intra-component couplings, $g_{1}$ and $g_{2}$, respectively. As we increase g, the dark component density exerts a repulsive forces on the bright component density and force it to localize in the center. As we pass the phase transition point when $g >2.3$, the density of the bright soliton component continues to decrease and therefore its amplitude decreases too and the density of the dark soliton component increases at a slow rate compared to the change in the bright component density. The difference between the rate of change in the density between the two components depends on their sizes. The dark soliton component is larger than the bright soliton component, as shown in Fig.~\ref{fig:FRHPRA:GS_many}, and therefore increasing the density of the dark soliton component will have a small effect on increasing its amplitude.

\begin{figure}[!htbp]
\centerline{\includegraphics[width=\columnwidth]{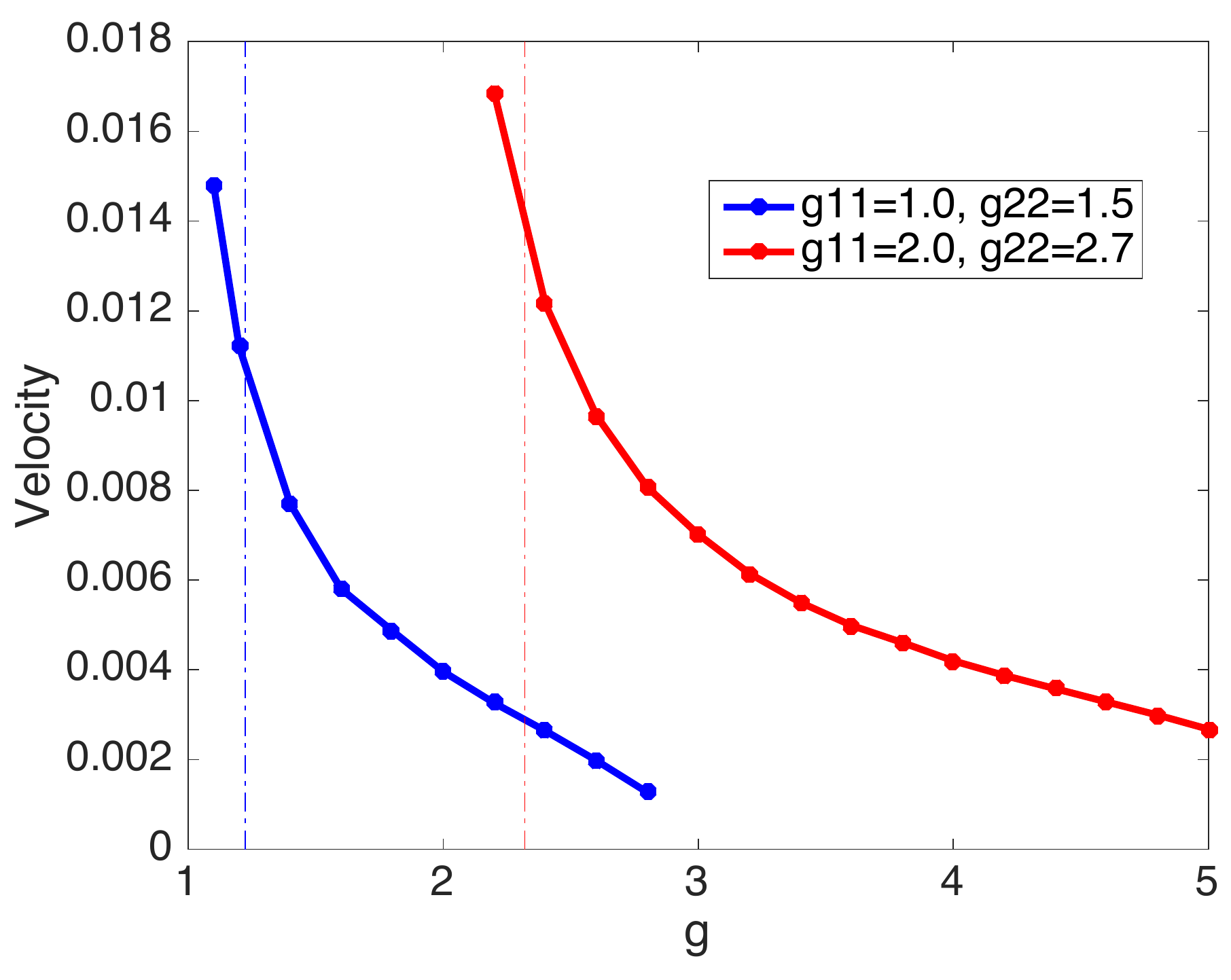}}
\caption{\label{fig:FRHPRA:g_vs_velocity}  \emph{Velocity of dark-bright soliton with phase imprint $\phi$=0.5 and different values of $g_{1}$ and $g_{2}$.} In the immiscible domain when $g >\sqrt{g_{1} g_{2}}$ we plot the velocity of the dark-bright soliton versus the coupling coefficient interaction, g, see Section~\ref{sec:FRHPRA:Units} for convert units. Note that a dark-bright soliton can be created as we get very close to  this line from the miscible domain. The amplitude of the bright soliton controls the rate of the velocity of the dark-bright soliton. As we increase the intercomponent coupling interaction, $g$, the amplitude of the bright soliton decreases as shown in Fig.~\ref{fig:FRHPRA:g_vs_A} and therefore the density of the bright soliton decreases too. Imprinting a phase on the small density bright soliton will have a small effect on dragging the dark soliton and therefore will result in a small velocity of the dark-bright soliton.}
\end{figure}

In Fig.~\ref{fig:FRHPRA:g_vs_velocity} the velocity of the dark-bright soliton drops quickly at the beginning then it slowly decreases as the coupling interaction increases. This behavior can be understood if we examine the density of the bright component. We find the form depicted in Fig.~\ref{fig:FRHPRA:g_vs_A}, i.e., that the amplitude (and therefor the density) of the bright component decreases as the coupling interaction increases. In this case, the imprinted phase on the “small” bright component will not pull the dark soliton quickly and therefore the velocity of the dark-bright soliton changes at a small rate as the bright component amplitude decreases.  In addition, the initial velocity of the dark-bright soliton when $g_{1}$=2.0 and $g_{2}$=2.7 is higher than the case when $g_{1}$=1.0 and $g_{2}$=1.5 because the difference between the amplitudes of the two components in the former case is less than in the latter. In other words, a phase imprinted on the bright component will have a bigger impact in the former case. The dashed lines distinguish the miscible and immiscible domains. Note that a dark-bright soliton can be created as we approach this line from the miscible domain.

\begin{figure}[!htbp]
\centerline{\includegraphics[width=\columnwidth]{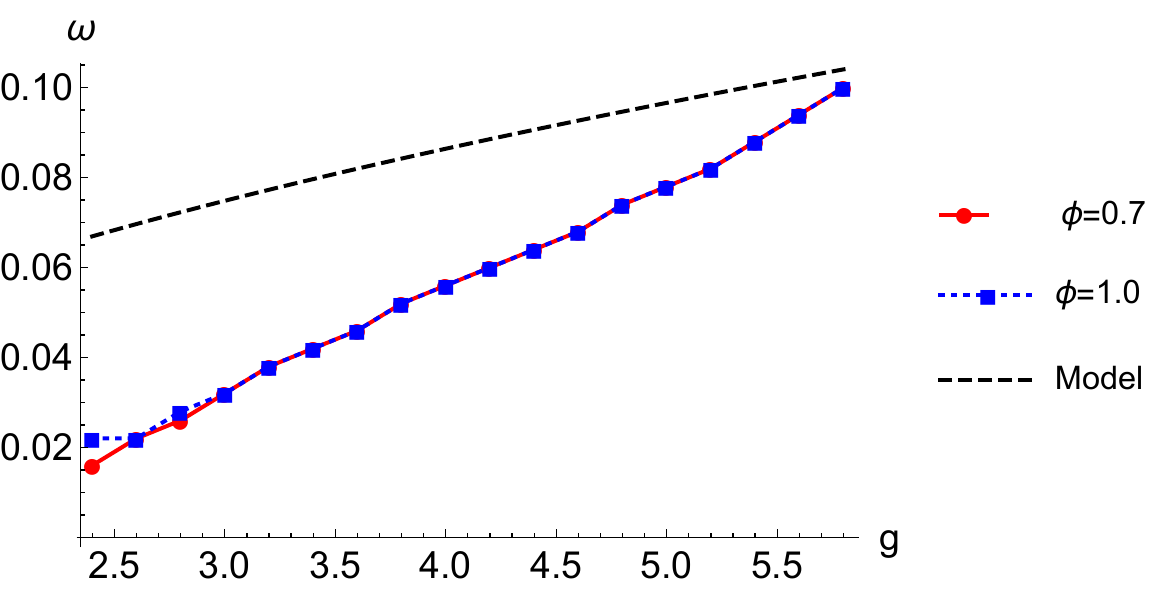}}
\caption{\label{fig:FRHPRA:g_vs_freq}\emph{Oscillation frequency of the two components  in the dark-bright soliton, with $g_{1}=2.0$, $g_{2}=2.7$ and $\phi=0.7$ and 1.0, obtained from numerically integrate Eq.~\eqref{eq:coupledNLSE} verses the oscillation frequency obtained from the analytical calculations, Eq.~\eqref{eq:oscillationfrequency}.} Numerically, the oscillation frequency of the two components versus the coupling coefficient $g$ for different values of $\phi$ shows that the oscillation frequency is amplitude independent in the case explored. We also plot the result from Eq.~\eqref{eq:oscillationfrequency} to compare the two outcomes from the analytical and numerical calculations.}
\end{figure}

In Fig.~\ref{fig:FRHPRA:g_vs_freq}, we first discuss the numerical results then we will discuss the comparison between these outcomes and the analytical results. Numerically, different values of imprinted phases on the bright component are shown in the figure ($\phi=0.7$ and $\phi=1.0$). The oscillation frequency of the two components versus the coupling interaction g is almost identical which means that the frequency is amplitude independent. Imprinted a large phase on the bright component can decouple the two components in the dark-bright soliton. In the case with $\phi=1.0$ the imprinted phase is large enough to cause a disturbance when the coupling coefficient is close to the miscible domain and therefore it shows a different oscillation frequency at the beginning. In the same figure we plot also the analytical results obtained from Eq.~\eqref{eq:oscillationfrequency}. We did not include the oscillation of the width i.e., the breather mode, in the analytical calculations because we can only perform the calculations for in-phase width oscillation analytically. In the numerical calculations the motion includes arbitrary-phase width oscillation and therefore including only the in-phase width oscillation in the analytical calculations will not have a significant improvement. The range of the values of g are bounded between two limits. In the lower limit, when $g<\sqrt{g_{1}g_{2}}$, that is in the miscible domain, the bright component in the DB exist on a top of a finite background (for example see Fig.~\ref{fig:FRHPRA:GS_many}). Therefore, imprinting phase on the bright soliton component to start the oscillation motion will also move the finite background density which will case a disturbance and affect the frequency results. The upper limit of the values of g are coming from the fact that for large g the ground state energy of the system does not support a DB soliton because of the strong intercomponent interactions between the dark component and the bright component

We see also in Fig.~\ref{fig:FRHPRA:g_vs_freq} that the comparison between the numerical and the analytical results become better as we increase the intercomponent interactions g. When g is close to the miscible domain the oscillation of the width of the two components is stronger due to the fact that g is small and therefore the width oscillation contribute to the oscillation of the two components. When g is large, the oscillation of the width of the two components become smaller due to the fact that the repulsive interaction between the two component is stronger and therefore it will force the two component to be confined in their region. Therefore as we increase g we will have a smaller contribution of the width oscillation mode in the oscillation of the two components which will improve the comparison between the numerical and the analytical results.


\begin{figure}[!htbp]
\center
\includegraphics[width=\columnwidth]{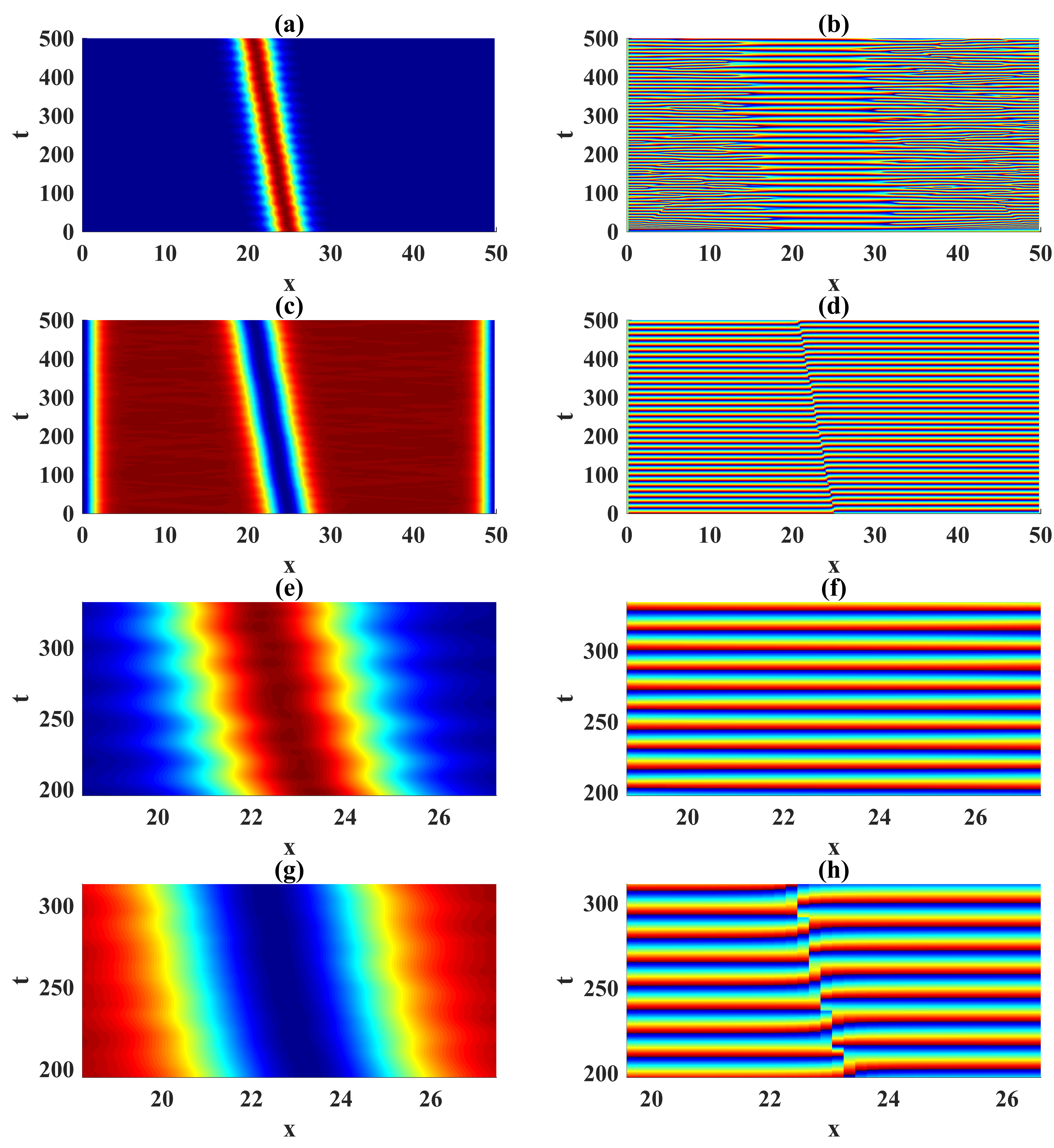}
\caption{\label{fig:FRHPRA:density_space} \emph{Oscillation of the two-component wave function $|u\left(x,t\right)|^2$ and $|v\left(x,t\right)|^2$ in the immiscible domain with $g_{1}=2.0$, $g_{2}=2.7$, $g=3.2$ and $\phi$=0.7.} In (a), (b), (c) and (d) represent the density and the phase of the bright and dark components, respectively. In figures (e), (f), (g) and (h) we plot the previous figures with a small time and space intervals to show the oscillations.}
\end{figure}

We obtained Fig.~\ref{fig:FRHPRA:density_space} by numerically integrate Eq.~\eqref{eq:coupledNLSE} using a pseudo-spectral method as mentioned in the introduction. Fig.~\ref{fig:FRHPRA:density_space} captures many features of the interactions of the two components in dark-bright soliton. For the oscillating bright component, Fig.(8a) and Fig.(8b) represent the density and the phase. For the oscillating dark component, Fig.(8c) and Fig.(8d) represent the density and the phase. The lower four panels show the same figures as the upper four panels with a zoom window on a small interval to display the oscillation. The interaction coefficients are $g_{1}$=2.0, $g_{2}$=2.7, $g$=3.2 and $\phi$=0.7. The oscillation frequency amplitude of the dark component decreases as we increase the interaction coefficient which in return make the observation of the oscillation in the dark component not obvious compared to the oscillation of the bright component as seen in Fig.~\ref{fig:FRHPRA:density_space}. For the above interaction coefficient values the amplitude of the bright component is almost half the amplitude of the dark component, as shown in Fig.~\ref{fig:FRHPRA:density_space} and Fig.~\ref{fig:FRHPRA:g_vs_A}.


\begin{figure}
\begin{subfigure}
  \centering
  \includegraphics[width=\columnwidth]{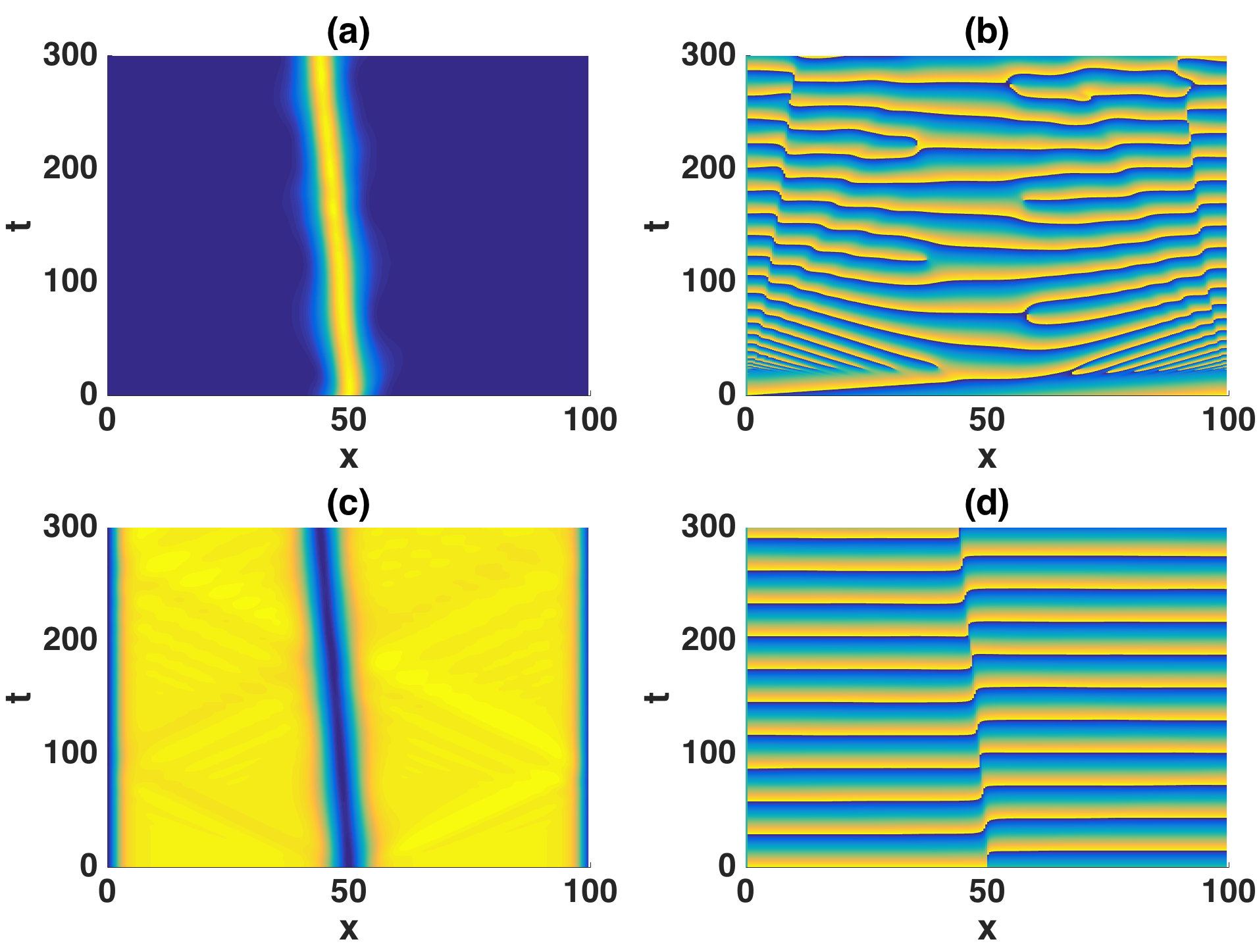}
\end{subfigure}%
\begin{subfigure}
  \centering
  \includegraphics[width=\columnwidth]{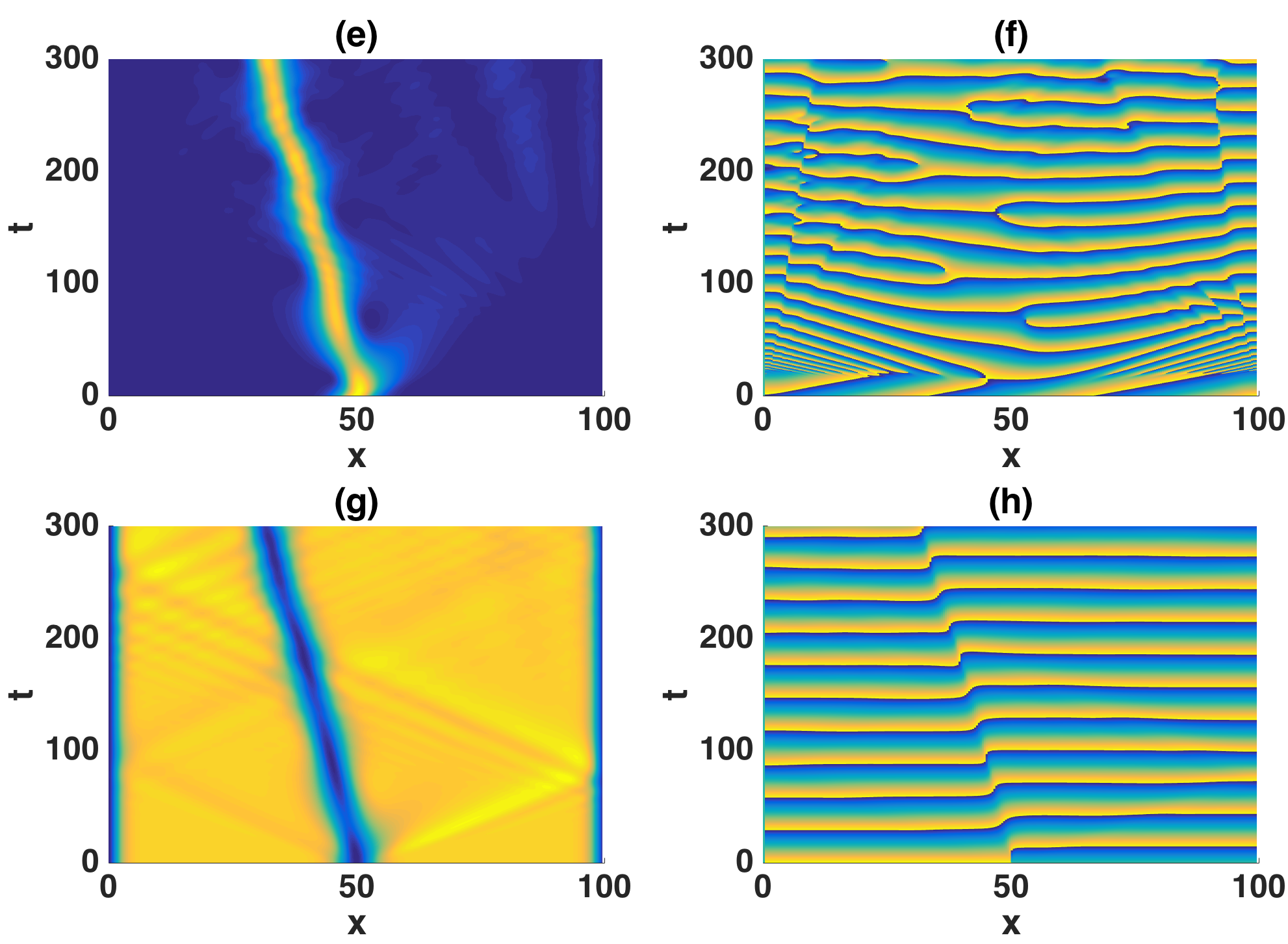}
\end{subfigure}%
\begin{subfigure}
  \centering
  \includegraphics[width=\columnwidth]{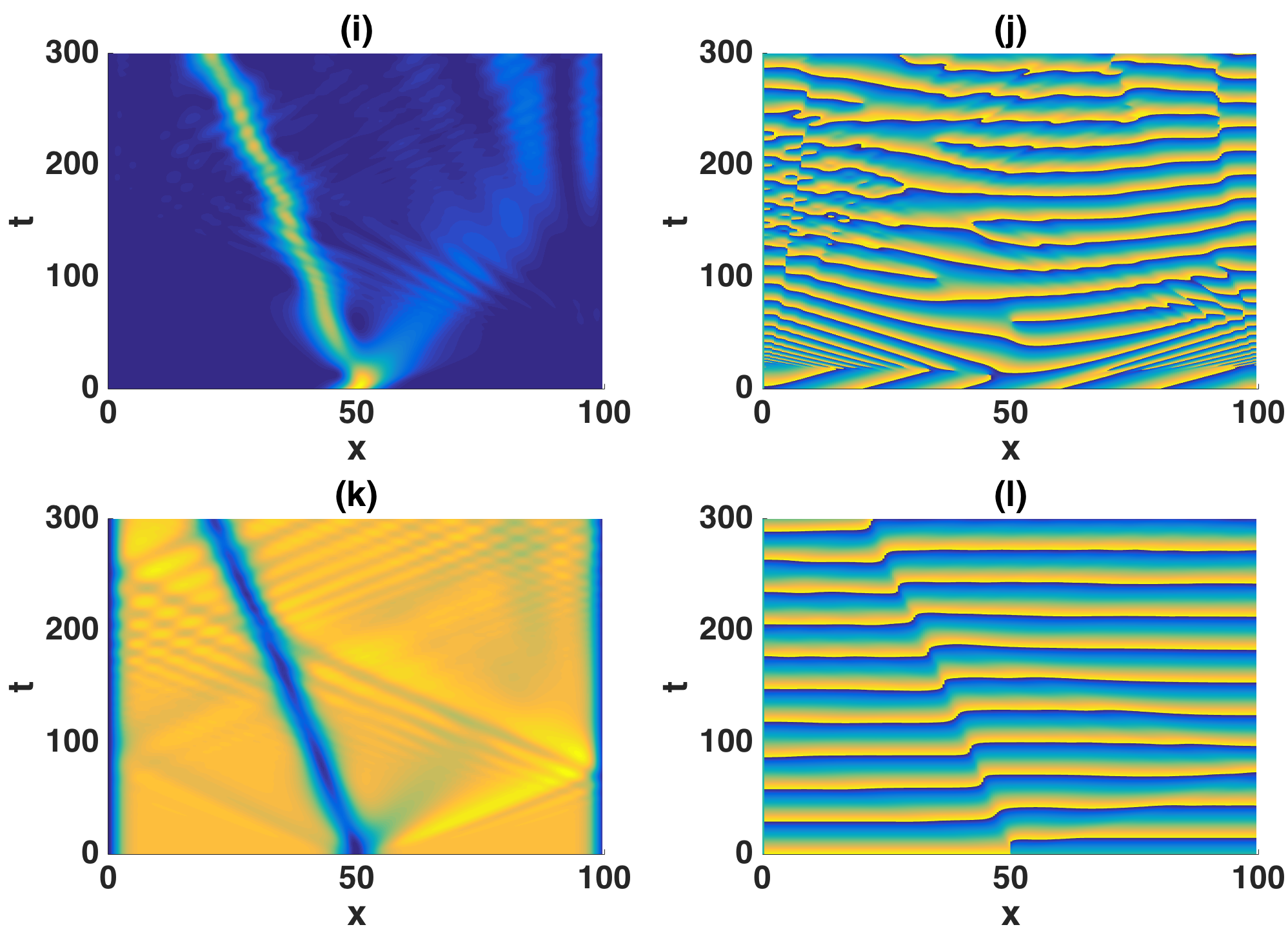}
\end{subfigure}
\caption{ \emph{Breaking the dark-bright by imprinting different values of the phase, $\phi$, on the bright component with interaction coefficients $g_{1}=2.0$, $g_{2}=2.7$, $g=2.6$.} In figures (a)-(d), (e)-(h) and (i)-(l) we have the values of the phase $\phi=$ 2, 6 and 10, respectively. In the left (right) panel, we have the density (phase) of the bright and the dark component, respectively.}
\label{fig:FRHPRA:breaking_VS}
\end{figure}



In Fig.~\ref{fig:FRHPRA:breaking_VS}, we plot the dark-bright soliton where we imprint a relatively large phase on the bright soliton component in order to unbind the dark-bright soliton. We emphasize that the bright component of a dark-bright soliton can only exist at long times in bound form.  When the imprinting phase is large (i.e. $\phi$ = 6 and 10) a significant portion of the bright soliton density escapes from the effective potential created by the dark soliton component and therefore breaks up the dark-bright vector soliton. Using the interaction coefficients mentioned in Fig.~\ref{fig:FRHPRA:breaking_VS} in Eq.~\eqref{eq:phi_1} in addition to setting $N_{1} = 0.503*10^5$, $N=1*10^5$,$\theta_{1}=1$,$\phi_{2}=1$,$\theta_{2}=2$ and the width=1 we find that the system oscillates as long as $\phi_{1}<3.4$ and the dark-bright vector soliton start to unbind above this value. We find this value in good agreement with the numerical results obtained in Fig.~\ref{fig:FRHPRA:breaking_VS} where we see that a significant fraction of the bright soliton component breaks away from the effective potential created by the dark soliton component around $\phi = 6$ and above.



\subsection{ Units}
\label{sec:FRHPRA:Units}

Typical experimental values for $^{87}$Rb BEC are $\omega_{\perp}\approx 2\pi\times 720$ Hz, $a_{s}\approx 5.1*10^{-9}$m and $N\approx 10^{5}$. For these parameters, the length scale is $\ell_{\bot} \approx 0.4\ \mu m$ and the time scale is $t_{\bot}\approx 0.22\ ms$.

\begin{table}[!htbp]
\centering
\begin{tabular}{||c|c|c|c||}
 \hline
 SI Units & Factor per Unit & Unitless & Unit \\ [0.5ex]
 \hline\hline
 $\tilde{x}$ & $0.4*10^{-6}$ & x & meter \\
 $\tilde{t}$ & $0.22*10^{-3}$ & t & second \\
 $\tilde{g_{ij}}$ & $13.7$ & $g_{ij}$ & $k_{B}$$\cdot$nK$\cdot$ $\mu$m \\
 $\tilde{\omega}$ & $4.5*10^{3}$ & $\omega$ & Hz \\
 $\tilde{u_{0}}^2$ & 33.9 & $u^2_{0}$ & $k_{B}$$\cdot$nK \\
 $\tilde{u}\left(\tilde{x},\tilde{t}\right)$ & $1.57*10^{3}$ & $u\left(x,t\right)$ & $\frac{1}{\sqrt{\mathrm{meter}}}$ \\
 $\tilde{v}\left(\tilde{x},\tilde{t}\right)$ & $1.57*10^{3}$ & $v\left(x,t\right)$ & $\frac{1}{\sqrt{\mathrm{meter}}}$ \\ [1ex]
 \hline
\end{tabular}
\caption{Converted Units.}
\label{table:1}
\end{table}
An example of using the units in the table to calculate the oscillation mode in $^{87}$Rb is by examine Fig.~\ref{fig:FRHPRA:g_vs_freq}. For g=4 we find that the oscillation frequency $\omega$ is 0.056. Using the units in Table~\ref{table:1} the equivalent SI units are $\omega$=252 Hz with g=$54.8$ $k_{B}$$\cdot$nK$\cdot$ $\mu$m which are reasonable numbers for an experiment in $^{87}$Rb.

\FloatBarrier

\section{Conclusions}
	\label{sec:FRHPRA:Conclusions}

We have calculated the normal modes of the system using hyperbolic secant for the bright component and hyperbolic tangent for the dark component. We found the velocity of each component depends on the phase imprinted following the known equation of the velocity of the condensate where the phase should depend on $x$ in order to move the dark-bright soliton components. In the dark component, the velocity also depends on the amplitude. There are two modes of the oscillation of the dark-bright soliton, the Goldstone mode, which we interpreted as a moving dark-bright soliton without internal oscillation of the two components, and the oscillation mode of the two components relative to each other. In addition, we have found numerically that in order to find a bright component in dark-bright soliton the density of the bright component is required to be less than the density of the dark component. This result was supported by analytical calculations in section~\ref{sec:FRHPRA:Normal Modes} where we found that in order to get oscillating bright and dark components we must meet the criteria mentioned above. In the numerical section, we have calculated different aspects of the interaction between the two components. Of particular interest is the two component oscillations in the dark-bright soliton where we found that the oscillation frequency does not depend on the phase imprinted which means that the frequency is amplitude independent. We showed the oscillation of the density and the phase of the two-component dark-bright soliton. Also, we have calculated the binding energy of the dark-bright soliton. We have compared the binding energy to the kinetic energy and the mean field energy of the dark-bright soliton in order to find the critical value of the imprinted phase on the bright component that breaks or unbinds the dark-bright soliton. Future work may extend the study to three-component solitons in different hyperfine states of the same condensate or for different species of atoms. In the multi-component case, the phase between the different components is coherent and the norm is not separately conserved. This material is based in part upon work supported by the NSF under grant number PHY-1306638.

\appendix
\section{Fixed point singularity}
\label{appendix:singularity}

Here we wish to prove that the system of Equations~\eqref{eq:EvolutionEqAll} does not posses a singularity. In particular, Eq.~\eqref{eq:8_d} with $l\equiv b(t)-d(t)$ become:

\begin{equation}
	\label{eq:A1}
	\begin{aligned}
\frac{d}{dt} \phi_{1}(t) =& \gamma \mathrm{csch}\left(\frac{l}{w} \right)^4 \left\{2 l
\left[ 2 +\mathrm{cosh}\left( 2 \frac{l}{w} \right) \right] \right.  \\ \nonumber  & \left. - 3w\mathrm{sinh}\left( 2 \frac{l}{w} \right) \right\} \\
=&  4l \gamma  \mathrm{csch}\left(\frac{l}{w}\right)^4 +2 l \gamma  \mathrm{cosh}\left(2\frac{ l}{w}\right)\mathrm{csch}\left(\frac{l}{w}\right)^4 \\ \nonumber &-3 w \gamma \mathrm{csch}\left(\frac{l}{w}\right)^4 \mathrm{sinh}\left(2\frac{l}{w}\right)
\end{aligned}
\end{equation}

When we expand the r.h.s of the above equation around the fixed point the terms $l^{-3}$ and $l^{-1}$ cancel out, and we are left with terms proportional to $l$. That is, the fixed point of the system (i.e. $l=0$) is valid. Note that we will not be able to address this fact if we work with Eq.~\eqref{eq:2ndODE} instead of Eqs.~\eqref{eq:EvolutionEqAll}.

\bibliographystyle{apsrev4-1}
\bibliography{library}
\end{document}